\documentstyle[NATO,numreferences]{CRCKAPB}

\begin{opening}
\title{New subdwarf B star periods}

\author{L. Morales-Rueda}
\author{T. R. Marsh}
\author{R. C. North}
\institute{Department of Physics \&\ Astronomy\\ University of
Southampton, UK}
\author{P. F. L. Maxted}
\institute{School of Chemistry \&\ Physics, Keele University, UK}

\end{opening}

\begin{document}

\section{Introduction}

Subdwarf B (sdB) stars are thought to be helium burning stars with
low mass hydrogen envelopes. Several evolutionary paths have been
proposed to explain the formation of these systems. One of these
scenarios is the evolution of the sdB progenitor within a binary
system. In fact \cite{m01} found that out of a sample of 36 sdBs, 21
of them reside in close binary systems. This result combined with
conclusions reached by other authors \cite{gls00} suggest that
two-thirds of sdBs are in binary systems.

With this in mind we have looked systematically at bright sdB stars
from the PG survey. By taking spectra at several different epochs we
have measured the radial velocity shifts caused by the motion of the
sdB star within the binary. Our data have been taken over a long time
base line (2 years) which allowed us to find longer period binaries
than known before. Here we present results for 29 sdB systems. The
methods we used to measure the radial velocities, to fit the radial
velocity data and to select the best alias are described in detail in
\cite{l02}.

\section{Orbital periods known up to now}

\begin{figure*}
\begin{picture}(100,0)(10,20)
\put(0,0){\includegraphics{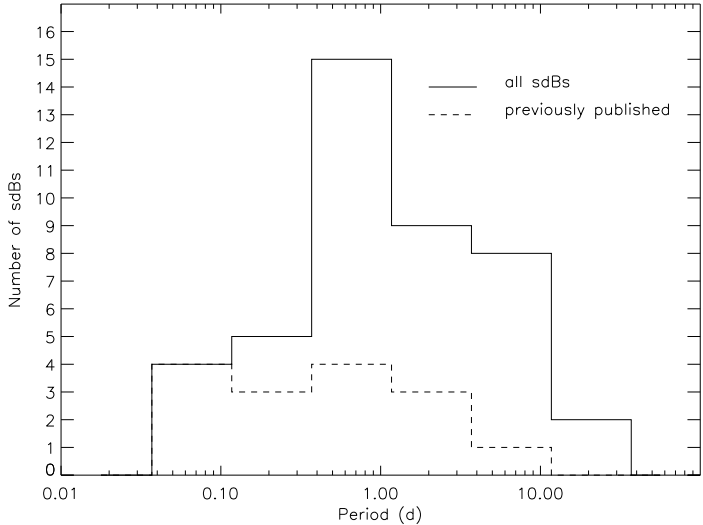}}
\noindent
\end{picture}
\vspace{65mm}
\caption{The solid line represents the number of sdBs known in binaries
  at a particular orbital period. The dashed line represents the sdBs
  binaries known previous to this study. Note that there seems to be
  and excess of systems at orbital periods of the order of one day.}
\label{fig1}
\end{figure*}

\begin{table*}
\begin{center}
\caption{List of the orbital periods measured for the 29 sdBs
  studied. T$_{0}$, the systemic velocity, $\gamma$, the radial
  velocity semi-amplitude, K, and the probability of the orbital
  period being further than 10\%\ from our favoured alias are also
  presented. The numbers quoted under the 10\%\ heading are actually
  the logs (base 10) of the probabilities.}

\begin{tabular}{lllccll}
Object & HJD (T$_{0}$) & Period (d)& $\gamma$ (km/s) & K (km/s) & 10\%\ & Ref. \\
 & $-$2450000 & & & & \\
\hline

KPD0025+5402 & 2159.386(9) & 3.571(1) & $-$7.8$\pm$0.7
& 40.2$\pm$1.1 & $-$2.3 & \cite{l02}\\

PG0133+114 & 2158.682(2) & 1.2382(2) & 6.0$\pm$1.0 & 
83.2$\pm$0.8 &$-$15.8 & \cite{l02}\\

PG0839+399 & 1914.06(6) & 5.622(2) & 23.2$\pm$1.1 &
33.6$\pm$1.5 & $-$3.7 & \cite{l02}\\

PG0849+319 & 1841.992(3) & 0.74507(1) & 64.0$\pm$1.5 &
66.3$\pm$2.1 & $-$4.2 & \cite{l02}\\

PG0850+170 & 1834.3(2) & 27.81(5) & 32.2$\pm$2.8 &
33.5$\pm$3.1 & $-$3.4 &\cite{l02} \\

PG0907+123 & 1840.62(3) & 6.1163(6) & 56.3$\pm$1.1 &
59.8$\pm$0.9 & $-$5.0 &\cite{l02} \\

PG0918+029 & 1842.310(4) & 0.87679(2) &
104.4$\pm$1.7 & 80.0$\pm$2.6 & $-$9.2 & \cite{l02}\\

PG0934+186 & 2376.58(1) & 4.05(1) & 7.4$\pm$2.9 & 
60.2$\pm$2.0 & $-$4.8 &\\

PG1017$-$086 & 2036.3940(5) & 0.072994(3) & 
$-$9.1$\pm$1.3 & 51.0$\pm$1.7 & $-$37.1 & \cite{m02}\\

PG1032+406 & 1888.66(2) & 6.779(1) & 24.5$\pm$0.5 &
33.7$\pm$0.5 & $-$2.0 &\cite{l02} \\

PG1043+760 & 1842.4877(7) & 0.1201506(3) &
24.8$\pm$1.4 & 63.6$\pm$1.4 & $-$4.7 & \cite{l02}\\

PG1110+294 & 1840.49(3) & 9.415(2) & $-$15.2$\pm$0.9 &
58.7$\pm$1.2 & $-$6.7 & \cite{l02}\\

PG1116+301 & 1920.834(2) & 0.85621(3) &
$-$0.2$\pm$1.1 & 88.5$\pm$2.1 & $-$4.8 & \cite{l02}\\

PG1230+052 & 2378.640(4) & 0.8372(2) & $-$43.4$\pm$0.8 & 
41.5$\pm$1.3 & $-$38.1 & \\

PG1244+113 & 2019.39(4) & 5.7520(7) & 9.8$\pm$1.2 & 
55.6$\pm$1.8 & $-$7.9 & \\

PG1248+164 & 1959.853(4) & 0.73232(2) &
$-$16.2$\pm$1.3 & 61.8$\pm$1.1 & $-$4.9 & \cite{l02}\\

PG1300+279 & 1908.310(7) & 2.2593(1) & $-$3.1$\pm$0.9
& 62.8$\pm$1.6 & $-$5.7 & \cite{l02}\\

PG1329+159 & 1840.579(1) & 0.249699(2) &
$-$22.0$\pm$1.2 & 40.2$\pm$1.1 & $-$3.6 & \cite{l02}\\

PG1512+244 & 1868.521(2) & 1.26978(2) &
$-$2.9$\pm$1.0 & 92.7$\pm$1.5 & $-$7.3 & \cite{l02}\\

PG1519+640 & 2391.888(2) & 0.539(2) & 1.5$\pm$0.6 & 
35.8$\pm$0.8 & $-$9.7 & \\

PG1528+104 & 2395.054(1) & 0.331(1) & $-$49.9$\pm$0.8 &
52.7$\pm$1.3 & $-$6.5 &\\

PG1619+522 & 1837.0(1) & 15.357(8) & $-$52.5$\pm$1.1 &
35.2$\pm$1.1 & $-$5.2 & \cite{l02}\\

PG1627+017 & 2001.267(1) & 0.829226(8) &
$-$43.7$\pm$0.5 & 73.6$\pm$0.9 & $-$68.9 & \cite{l02}\\

PG1716+426 & 1915.806(5) & 1.77732(5) &
$-$3.9$\pm$0.8 & 70.8$\pm$1.0 & $-$5.2 & \cite{l02}\\

PG1725+252 & 1901.3977(8) & 0.601507(3) &
$-$60.0$\pm$0.6 & 104.5$\pm$0.7 & $-$124.4 & \cite{l02}\\

PG1743+477 & 1921.1183(7) & 0.515561(2) &
$-$65.8$\pm$0.8 & 121.4$\pm$1.0 & $-$26.5 & \cite{l02}\\

HD171858 & 2132.241(6) & 1.529(8) & 73.8$\pm$0.8 &
93.6$\pm$0.7 & $-$6.9 & \cite{l02}\\

KPD1946+4340 & 2159.0675(5) & 0.403739(8) &
$-$5.5$\pm$1.0 & 167.0$\pm$2.4 & $-$15.3 & \cite{l02}\\

KPD2040+3955 & 2288.465(5) & 1.48291(8) &
$-$11.5$\pm$1.0 & 95.1$\pm$1.7 & $-$2.9 & \\

\end{tabular}
\end{center}
\label{tab1}
\end{table*}

Fig.~\ref{fig1} presents a histogram with all the sdB periods known up
to now compared with the sdB periods known previous to this study. The
long time base line of our data has allowed us to detect systems with
orbital periods of the order of tens of days whereas previously there
seemed to be a tendency for sdBs to be present only in short period
binaries. One second important feature we see in this histogram is
that we do not find any gaps in the distribution of orbital periods
like we do with cataclysmic variables but there seems to be a tendency
for sdB binaries to have periods of the order of one day.
\cite{green02} find the same tendency in their own sample of sdB
binaries. Table~1 presents the orbital solutions found for the 29
systems discussed. A more detailed table including periods measured by
other authors is presented in \cite{l02}.

\section{Consequences for evolutionary theories}

\begin{figure*}
\begin{picture}(100,0)(10,20)
\put(0,0){\includegraphics{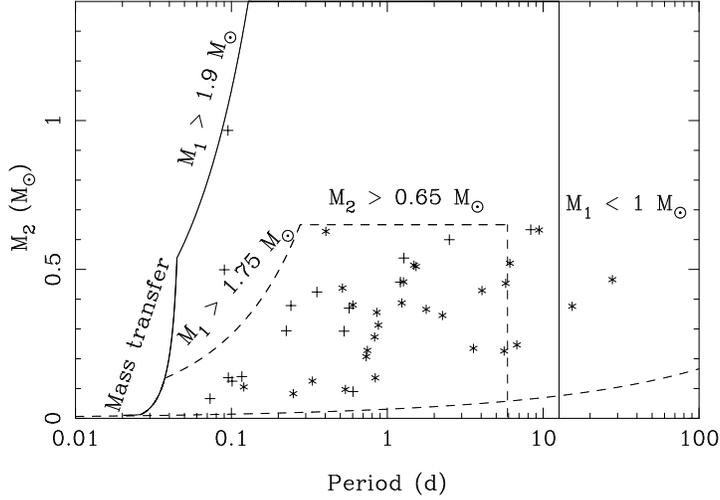}}
\noindent
\end{picture}
\vspace{68mm}
\caption{Minimum companion mass versus orbital period for
  sdBs. Asterisks represents systems studied in this paper, plus
  symbols represent systems previously published.}
\label{fig2}
\end{figure*}

By using the well known equation for the mass function of the
companion and assuming that the mass of the sdB is 0.5\,M$\odot$ we
obtain values for the minimum mass of the companions to sdBs.
Fig.~\ref{fig2} shows the distribution of minimum companion masses
versus orbital period for the 29 systems presented in this paper
(represented by asterisks) and the systems known previously
(represented by plus symbols). Also plotted in the graph (solid lines)
are the different model constraints that account for: i) the lack of
mass transfer in sdB binaries, ii) the lower limit for the mass of the
sdB progenitor to ensure that the helium flash occurs, iii) the
maximum mass for the sdB companion if it is not easily visible, iv)
the upper limit on the orbital period given that the mass of the sdB
progenitor had to be more than 1 M$_{\odot}$ to evolve to this stage.
The bottom dashed line indicates the minimum companion mass we can
detect due to the resolution of our data. Tighter constrains (dashed
lines), that contain most of the systems presented, are also plotted.
These constraints have been calculated assuming an envelope ejection
efficiency $\alpha_{\rm CE}\lambda$ = 0.5 and a metallicity [Fe/H] =
$-$0.6. If we consider higher envelope ejection efficiencies or higher
metallicities, all the constraints move towards the right of the graph
explaining the presence of the high period systems. In this second
case though, short period systems with high mass companions are not
explained by our model.

\section{Conclusions}

We find that most of the systems with orbits known up to now show a
range of parameters (periods and masses of the companions to the sdB
stars) that can be explained with an evolutionary model that consists
of the formation of a common envelope after the onset of mass transfer
when the sdB progenitor was at the tip of the red giant branch. Large
orbital period systems can be explained by assuming either high common
envelope ejection efficiencies or large metallicities for the sdB
progenitor. But we find that these models cannot explain
simultaneously large period systems and short period systems with
large mass companions like KPD1930+2752 \cite{mmn00}.

\end{document}